\documentclass[final,5p,times,twocolumn]{elsarticle}

\usepackage{graphicx}

\usepackage{amsmath}
\usepackage{amssymb}

\usepackage[font=footnotesize,labelfont=bf]{caption}
\usepackage[font=footnotesize,labelfont=bf]{subcaption}

\usepackage{booktabs}
\usepackage{multirow}

\journal{Computer Standards \& Interfaces}

\begin{document}

\begin{frontmatter}



\title{Standards for enabling heterogeneous IaaS cloud federations}


\author{{\'A}lvaro L{\'o}pez Garc{\'i}a\corref{cor1}}

\cortext[cor1]{Corresponding author}
\ead{aloga@ifca.unican.es}

\author{Enol Fern{\'a}ndez del Castillo}
\ead{enolfc@ifca.unican.es}

\author{Pablo Orviz Fern{\'a}ndez}
\ead{orviz@ifca.unican.es}

\address{Institute of Physics of Cantabria, Spanish National Research Council --- IFCA (CSIC---UC).\\
Avda. los Castros s/n. 39005 Santander, Spain}

\begin{abstract}
    As different Cloud Management Frameworks and resource providers are
    settling in the market there is the need to grasp the interoperability
    problems present in current infrastructures, and study how existing and
    emerging standards could enhance user experience in the cloud ecosystem.

    In this paper we will review the current open challenges in Infrastructure
    as a Service cloud interoperability and federation, as well as point to the
    potential standards that should alleviate these problems.
\end{abstract}


\end{frontmatter}

\section{Introduction}
\label{sec:introduction}

Cloud computing is still considered as an emerging technology, although it is
now leaving its infancy phase. Standardization in the cloud was not considered
as an urgent topic by the industry \cite{Dillon2010}, as it is often associated
to rigidity, not leaving much room for the innovation needed on the early
stages of the technology \cite{Tassey2000}.

Over the last years, a large number of commercial cloud providers have emerged
in the market. Each of those vendors tries to differentiate their
infrastructure from their competitors offering added value features on their
resources. This has led to a situation where several closed and proprietary
interfaces have evolved over the time, some being claimed as \emph{de-facto}
standards by the industry. The resulting scenario consists on infrastructures
using different solutions that are actually incompatible and not interoperable,
locking users inside a single provider. These vendor lock-ins are often
considered a desirable feature by commercial providers, as a way of keeping
users attached to their resources and services, but it is perceived negatively
by cloud users and customers \cite{Borenstein2011}.


More recently, several open source Cloud Management Frameworks (CMFs) have
appeared in the cloud ecosystem. Some of them decided to adopt the most popular
commercial and proprietary interface, implement a compatibility layer that
tries to deliver the same functionality; whereas others have built their own
interface. Both decisions are contributing to adding more entropy and
heterogeneity into the cloud ecosystem. Users willing to exploit several
infrastructures face a discouraging panorama, with strong industrial actors
driving the developments that have promoted a situation where proprietary and
industry-driven interfaces and protocols have dominated the cloud landscape for
years \cite{Petcu2013}.

As the cloud computing paradigm is maturing and its heterogeneity is growing,
cloud interoperability and federation are becoming areas of concern
\cite{Badger2011, James2015}.  Federation and interoperability are nowadays
considered as one of the main pressing issues towards cloud computing adoption
\cite{Schubert2010}. The vendor lock-ins that currently exist are perceived
negatively by users, therefore building and defining frameworks for cloud
interoperability is becoming therefore a topic that is gaining more and more
interest
\cite{Campos2012,Bernstein2009,parak2014challenges,Feldhaus2014,FernandezdelCastillo2015}.
Moreover, political and government bodies such as the European Commission have
stated their position towards the promotion of Open Standards for ensuring
interoperability in clouds for science and public administration
\cite{Idabc2004,Cattaneo2013}.

Nevertheless, cloud federation goes beyond just making several clouds
interoperable \cite{Kurze2011}. A federation should enable the collaboration
and cooperation of different providers in delivering resources to the users
when a single resource provider is not able to satisfy the user demands, in a
collaborative way. Therefore, on top of the interoperability and portability
issues, there are several challenges that any federation must tackle.

In this paper we will review the open challenges when building an interoperable
cloud federation. We will review the existing enabling standards that can be
used to leverage the construction of such a federation of Infrastructure as a
Service (IaaS) providers based on open standards. We will focus on an
\emph{horizontal federation} between different IaaS providers, therefore a
\emph{vertical federation} spanning several layers is out of the scope of this paper.

In Section~\ref{sec:related} we will present the related work in the area.
In Section~\ref{sec:challenges} we will present the biggest
challenges that an interoperable cloud federation must assess.
In Section~\ref{sec:standards} we focus on the existing and raising standards and
how they can be used to tackle the problems presented in
Section~\ref{sec:challenges}.
Finally we
present our conclusions in Section~\ref{sec:conclusions}.

\section{Related work}
\label{sec:related}

Some work and research has been done into cloud interoperability, although a
lot of the work is regarding cloud \emph{portability} between different cloud
infrastructures.

There are many non academic works regarding the need or lack thereof for a
\emph{Cloud standard}. However, authors agree that there would be not such a
unique standard to rule all the cloud aspects. Some preliminary work regarding
the need of standards for the cloud has been done in the past
\cite{Borenstein2011,Machado2009}.

The United States' National Institute of Standards and Technology has surveyed
the existing standards for interoperability, performance, portability, security
and accessibility in the NIST Cloud Computing Standards Roadmap
\cite{Bumpus2013}. However, there are some aspects like information discovery
or accounting that are missing in this study.

G. Lewis \cite{LewisTheRole2012} report tackles several standardization areas
such as workload management, data and cloud management APIs, concluding that
there will be not a single standard for the cloud due to pressures and the
influences of existing vendors.  The author states that an agreement on a set
of standards for each of the needed areas would reduce the migration efforts
and enable the third generation of cloud systems.

Harsh et al. \cite{Harsh2012} work surveyed the existing standards for the
management of cloud computing services and infrastructure  within the Contrail
project so as to avoid vendor lock-in issues and ensure interoperability.  In
the same line, Zhang et al. \cite{Zhang2013} performed a quite complete survey
regarding Infrastructure as a Service access, management and interoperability,
studying OVF, CDMI and OCCI \cite{Zhang2013}. However, they have not entered
into other details and challenges such as accounting or information discovery.

On top of those academic efforts, some open source Cloud Management Frameworks
(CMFs) have started to take into consideration the federation issues. There are
development efforts aimed to make possible to federate different aspects of
distributed cloud infrastructures to an extent:

\begin{itemize}
    \item OpenStack \cite{web:openstack} implements several levels of
        federation by the usage of cells and regions. The former allows to run
        a distributed cloud sharing the same API endpoint, whereas the latter is
        based on having separate API endpoints, federating some common
        services: OpenStack also allows the usage of a federated authentication
        mechanism \cite{Chadwick2014}, so that the identity service is able to
        authenticate users coming from trusted external services or from
        another identity service.
    \item CloudStack \cite{web:cloudstack} follows the same line and implements
        the concept of regions in their software.
    \item OpenNebula \cite{web:opennebula} makes possible to configure several
        installations into a tightly integrated federation, sharing the same
        users, groups and configurations along several installations.
    \item Eucalyptus \cite{web:eucalyptus} provides with identity and
        credential federation.
\end{itemize}

However, all of them rely on the fact of federating several instances of the
same software stack (i.e. several OpenNebula installations, for instance),
being impossible or difficult to federate disparate and heterogeneous
infrastructures (e.g.  an OpenStack installation together with an OpenNebula
instance).

On top of that, there a few prominent existing federated infrastructures, some
of them being built on top of standards, others not. Some examples of
standards-based federations are the EUBrazil Cloud Connect \cite{web:eubrazil},
whose middleware is being based on standards for interoperability
\cite{DiasdeLima2014}; and the European Grid Initiative (EGI) \cite{web:egi},
that started as a federation of grid sites, took the strategic position of
exploring and adopting a technology agnostic and based on open standards cloud
\cite{FernandezdelCastillo2015} into their services portfolio. In this context,
the Open Science Cloud initiative \cite{Koski2015} has outlined that
interoperable, distributed and open principles should drive the evolution of
Science Clouds as the key to success.

\section{Cloud Federation Open Challenges}
\label{sec:challenges}

As we briefly exposed in Section~\ref{sec:introduction}, a cloud federation
should take into account other aspects apart from interoperability and
portability such as authentication, authorization or accounting. In the
following sections we will elaborate on the open challenges regarding cloud
federation.

\subsection{On Uniform Access and Management}

One of the first obstacles that a heterogeneous cloud federation has to
overcome is the lack of a unified cloud interface. Evolving from commercial
cloud providers, each middleware implements their own ---proprietary or not---
interface. Some open CMFs implement an Amazon Web Services (AWS) EC2
\cite{web:aws} compatibility layer, since it was considered as the most popular
commercial interface for the cloud.

The adoption of the AWS EC2 API could make two different CMFs being
interoperable, but it presents several obvious drawbacks. First of all, its
usage and promotion introduces a vendor lock-in, as users can be locked into
one infrastructure if the original vendor decides to change its API from one
day to another. A proprietary API is subject to change without prior advice by
the original vendor. This will render into incompatibilities between providers
and CMFs other than the original creator of the API, Amazon in this case.
Implementers of such proprietary interfaces need to keep aligned with the
reference implementation, and are forced to invest time in following the
modifications so that they ensure that its implementation remains compatible.

Secondly, the EC2 Query API is not RESTful. Even if it uses the standard
components of the HTTP protocol to represent API actions it does not use the
HTTP message components to indicate the API operations, being them expressed as
parameters (in the URI parameters of a GET request or in the body of a POST
request). This URI-based parameter passing is not enough for defining an
interoperable API allowing an standardized implementation. Moreover, non being
RESTful introduces additional complexity for developers to create applications
that exploit it, as they have to learn the semantics being used instead of the
well known REST architectural style.  Lastly, the usage of the query component
of an URI to obtain hierarchical data goes against the RFC-3986 "Uniform
Resource Identifier (URI): Generic Syntax" \cite{Berners-Lee2014}, as it states
that "The query component contains non-hierarchical data that along with data
in the path component, serves to identify a resource (...)".

\subsection{On Portability}

Cloud computing leverages virtualization technologies to abstract the resources
being offered to the users. Several virtualization hypervisors (such as Xen,
KVM, VMWare, Hyper-V) exist in the market, and each cloud provider will be
using the one of its choice. Moreover, recently operating system level
virtualization (that is, container-based such as LXC, OpenVZ and Docker)
have entered the game and they are being more and more adopted by the
providers.

This situation renders difficult the migration of one virtual appliance
prepared to be executed in one cloud provider using one virtualization
technology to another provider with a different underlying technology.
Moreover, the underlying technology is hidden and abstracted from the users by
the CMFs, hence even if they had the technological skills to prepare and modify
a virtual appliance to be executed on another hypervisor they would have found
difficulties in doing so. Porting one Virtual Machine to another hypervisor may
require access to consoles and debugging output that cloud providers may be
reluctant to provide.

\subsection{On authentication and authorization}

The delivery of an homogeneous authentication and authorization via a federated
identity management system in distributed environment is a challenging topic
\cite{Chadwick2014,Shim2001} not exclusive to cloud computing. As a matter of
fact, cloud computing is just another player in the game. Such system should
facilitate flexible authentication methods and federated authorization
management.

The lack of a federated identity management systems makes difficult to
manage the users globally at the federation, that is, specifying what resources
a user is able to access or globally disabling a user becomes a challenging
task.

Moreover, this challenge has two additional faces  as it affects both the users
and the resource providers.

\begin{itemize}
    \item From the user's perspective, they are forced to cope with the burden
        of managing several credentials and identities. Moreover, client
        tools need to deal with them as well, identifying that \emph{identity
        A} should be used against \emph{provider A} but not \emph{provider B}.
    \item It increases the management complexity for the resource providers. If
        there is not such a federated identity management system, users are to
        be managed manually, thus incurring in a tremendous overhead.
\end{itemize}

\subsection{On information discovery}

Once a federation is established, the next challenge is how users are able to
discover what resources and capabilities are offered by the federation so that
they can consume them.

This information may be exposed to the users via each of the middleware native
APIs by each of the resource providers participating in the federation, but it
is not structured in an homogeneous way so that clients can fetch that
information and help users making a decision.

\subsection{On accounting and billing}

In federated infrastructures it is often required to keep track of resource
usage for each user and group at every individual provider, so that this
information is shared and aggregated at the federation level and users are
accounted properly. This aspect is tightly coupled with the federated identity
management systems, as users need to be unambiguously identified throughout the
infrastructure.  Currently, each Cloud Management Framework and provider may
have their own account method, but there is no common way for accessing and/or
aggregating that information at the federation level.

\section{Federation enabling standards}
\label{sec:standards}

It may be possible to obtain interoperability without the usage of Open
Standards \cite{web:open_standards}, but it is arguably a more logical
way to develop an interoperable federation based on them.

There is not a unique cloud standard to rule all of the aspects regarding
clouds \cite{Machado2009}, neither there is such a federation standard.
Nevertheless, there are several well established standards covering some of the
open issues described in Section~\ref{sec:challenges}, developed prior to the
raise of Cloud computing and that can be simply reused, adapted or updated to
fill in the needed gaps. On top of them there is a number of emerging standards
being developed specifically to cover more specific cloud computing topics.

It is the combination of both ---existing and emerging standards--- they key to
solve the federation and interoperability issues described in
Section~\ref{sec:challenges}, as we will describe through the rest of the
section.

\subsection{Uniform access and management}

Several organizations and standardization bodies have started working from the
early stages of cloud computing trying to build standards for cloud
management. Currently, the most prominent examples regarding IaaS computing
and storage management are OCCI, CIMI, TOSCA and CDMI:

\begin{description}
    \item[OCCI] The Open Grid Forum (OGF) \cite{web:ogf} has proposed the Open
        Cloud Computing Interface (OCCI)
        \cite{Metsch2011,Nyren2010,Metsch2010}, focusing on facilitating an
        interoperable access and management of IaaS cloud resources. OCCI
        offers different renderings over the HTTP protocol, leading to a
        RESTful API implementation.

    \item[CIMI] The Cloud Infrastructure Management Interface (CIMI)
        \cite{davis2012cloud} is a
        proposal from the Distributed Management Task Force (DMTF)
        \cite{web:dmtf} that has
        been recently registered as an ISO/IEC standard \cite{CIMIIso}. CIMI
        targets the management of the life-cycle of the IaaS resources, offering
        a RESTful API over the HTTP protocol with various renderings.

    \item[TOSCA] The Topology and Orchestration Specification for Cloud
        Applications (TOSCA) \cite{Lipton2013} is an standard from the
        Organization for the Advancement of Structured Information Standards
        (OASIS) \cite{web:oasis}. TOSCA provides a language to describe
        composite services and applications on a cloud, as well as their
        relationships (i.e. the topology), makes also possible to describe its
        operational and management aspects (i.e. its orchestration). Although
        TOSCA is at a higher level than simply managing the IaaS resources
        ---it is more focused on the orchestration of the resources---, it
        should be considered as a complementary standard for the management of
        the resources.

    \item[CDMI] The Storage Networking Industry Association (SNIA) has proposed
        the Cloud Data Management Interface (CDMI), defining an interface to
        perform different operations (creation, retrieval, update and removal)
        on data stored on a cloud.
\end{description}

\subsection{Portability}

The Open Virtualization Format (OVF) \cite{crosby2009open} is a standard
developed by the DMTF for packaging and describing a Virtual Appliance (VA),
comprised of an arbitrary number of virtual machines in a portable and vendor
neutral format. An OVF package contains a XML description (e.g. hardware
configuration, disks used, network configuration, contextualization
information, etc.) of each component of the VA.

\subsection{Authentication and authorization}

There is a large number of standards that can be used for authentication and
authorization. The implementation and adoption of one technology or another
will eventually depend on the infrastructures that are going to be federated,
and there will be no silver bullet that will fit all of the existing
infrastructures. Authentication and Authorization sometimes imply political
aspects that are out of the scope of the standardization efforts.

The X.509 Public Key Infrastructure \cite{rfc2459} has been used for
authentication in the grid world via the Grid Security Infrastructure (GSI),
based on X.509 certificate proxies \cite{rfc3820}. Authorization is done by
embedding Attribute Certificates (AC) into the proxy, containing assertions
about the user. The most notable service is the Virtual Organization
Management System (VOMS) \cite{Venturi2007}, being used in some federated
cloud infrastructures \cite{LopezGarcia2013}. However, X.509 certificates are
not considered being user-friendly in spite of being settled on several
distributed infrastructures over the years.

The OASIS Security Assertion Markup Language (SAML) \cite{oasis2005protocols}
is built in X.509 and defines a way to define authentication and attribute
assertions in XML. Shibboleth \cite{shibboleth} is an implementation of SAML
and is focused on the federation of resource providers with different
authentication and authorization schemes.  Several projects have started
looking at SAML and Shibboleth \cite{qiang2012standards,murri2011gridcertlib} as
a promising way to provide access to distributed infrastructures, although they
have not substituted X.509 yet.

OAuth 1.0 \cite{hardt2012oauth} and 2.0 \cite{oauth} is an IETF open standard
for authorization, providing delegated access to some resources on behalf of
the resource owner. OAuth has not been designed for authentication, therefore
OpenID Connect (OIDC) \cite{sakimura2014openid} has been developed as an
authentication layer on top of OAuth 2.0.

\subsection{Information discovery}

Information discovery is a problem present in other federated computing
paradigms such as Grids. The Grid Laboratory Uniform Environment (GLUE) Schema
---in its versions 1.x \cite{Andreozzi2007} and 2.0 \cite{Andreozzi2009}--- has
been designed by the OGF in order to create an information model relying on the
knowledge and experience from the operations of several large Grid
infrastructures.

The current GLUE 2.0 specification \cite{Andreozzi2009} only defines a
conceptual model. It makes possible to publish, separately from the standard,
concrete data model profiles that will dictate how the information is generated
and used for in concrete implementation, infrastructure, etc. Therefore, the
OGF GLUE 2.0 schema is a good candidate for publishing information relative to
cloud infrastructures.

\subsection{Accounting}

As with the information discovery, the accounting problem is a problem that has
been already tackled in the grid. The OGF Usage Record (UR) 2.0
\cite{Cristofori2013} defines a common format to share and exchange basic
accounting data, coming from different providers and different resources. It
supersedes and integrates the different resource usage records that leveraged
the previous UR 1.0 in the various infrastructures that implemented it.

The OGF UR does not specify how the records should be exploited (e.g. how they
should be exported, used, aggregated, summarized, etc.) or transported.
Examples on how the UR is used exist in projects such as RESERVOIR
\cite{Elmroth2009} and infrastructures such as EGI \cite{web:egi}.

\section{Conclusions}
\label{sec:conclusions}

As we explained, cloud federation involves a lot of different areas and
challenges ---management, authentication, accounting, interoperability, etc---,
therefore there is not a unique standard for it. However, there is a set of
settled and emerging standards that can cover all the federation aspects and
their problematics, as summarized in Table~\ref{tab:standards}.

\begin{table}[h!]
    \centering
    \begin{tabular}{ll}
        Challenge & Enabling Standard \\
        \toprule
        \multirow{3}{*}{Uniform access and management} & OGF OCCI \cite{Metsch2011,Nyren2010,Metsch2010}\\
                                                       & DMTF CIMI \cite{davis2012cloud}\\
                                                       & OASIS TOSCA \cite{Lipton2013}\\
        \cmidrule(l{0.85em}r{0.85em}){1-2}
        Portability & DMTF OVF \cite{crosby2009open}\\
        \cmidrule(l{0.85em}r{0.85em}){1-2}
        \multirow{3}{*}{Authentication and authorization} & OASIS SAML \cite{oasis2005protocols}\\
                                                          & OpenID Connect \cite{sakimura2014openid} \\
                                                          & X.509 \cite{rfc2459}\\
        \cmidrule(l{0.85em}r{0.85em}){1-2}
        Information discovery & OGF GLUE \cite{Andreozzi2007,Andreozzi2009} \\
        \cmidrule(l{0.85em}r{0.85em}){1-2}
        Accounting & OGF UR \cite{Cristofori2013} \\
        \bottomrule
    \end{tabular}
    \caption{Summary of enabling standards}
    \label{tab:standards}
\end{table}

In this document we have presented the existing challenges that need to be
tackled for building an interoperable federation of cloud providers and we
have surveyed the existing and arising standards that can be used to solve those
problems.  Current Cloud Management Frameworks should adopt these existing
standards for the functionality they are offering, so as to avoid vendor
lock-in issues and to ensure a that proper interoperability is delivered to the
users.

The European Commission is encouraging the usage of Open Standards in its
''European Interoperability Framework for pan-European eGovernment Services``
\cite{Idabc2004}. Similarly, the United Kingdom Government provided a similar
set of principles, adopted 2014 \cite{web:open_standards:uk}. Other European
initiatives, such as the Open Science Cloud \cite{Koski2015}, are also
promoting the usage of Open Standards. Cloud federations must take into account
these recommendations and they should promote its usage as the path to a
successful federation and interoperability.

\section*{Acknowledgements}

The authors acknowledge the financial support from the European Commission (via
EGI-InSPIRE Grant Contract number RI-261323).

The authors also want to thank the IFCA Advanced Computing and e-Science Group.

\bibliographystyle{elsarticle-num}
\bibliography{references}

\end{document}